\documentclass[aip,jcp,twocolumn,superscriptaddress]{revtex4}
\usepackage{graphicx}
\usepackage{amsmath}
\usepackage{amssymb}
\usepackage{pifont}
\usepackage{color}

\begin{document}

\title{Disjoining Pressure of an Electrolyte Film Confined between Semipermeable Membranes}

\author{Salim R.~Maduar}
\affiliation{A.N.~Frumkin Institute of Physical
Chemistry and Electrochemistry, Russian Academy of Sciences, 31
Leninsky Prospect, 119991 Moscow, Russia}
\affiliation{Faculty of Physics, M.V.~Lomonosov Moscow State University, 119991 Moscow, Russia }

\author{Olga I.~Vinogradova}

\affiliation{A.N.~Frumkin Institute of Physical
Chemistry and Electrochemistry, Russian Academy of Sciences, 31
Leninsky Prospect, 119991 Moscow, Russia}
\affiliation{Faculty of Physics, M.V.~Lomonosov Moscow State University, 119991 Moscow, Russia }
\affiliation{DWI - Leibniz Institute for Interactive Materials,  RWTH Aachen, Forckenbeckstr. 50, 52056 Aachen, Germany}

\date{\today}
\begin{abstract}

We consider an electrolyte solution confined by infinitesimally thin semipermeable membranes in contact with a salt-free solvent. Membranes are uncharged, but since small counter-ions  leak-out into infinite salt-free reservoirs, we observe a distance-dependent membrane potential, which generates a repulsive electrostatic disjoining pressure. We obtain the distribution of the potential and of ions, and derive explicit formulas for the disjoining pressure, which are validated by computer simulations.  We predict a strong short-range power-law repulsion, and a weaker long-range exponential decay. Our results also demonstrate that an interaction between membranes does strongly depend on the screening lengths, valency of an electrolyte solution, and an inter-membrane film thickness.
Finally, our analysis can be
directly extended to the study of more complex situations and some biological problems.

\end{abstract}

 \maketitle

\section{Introduction}

Ion or Donnan equilibria which exist in the presence of \emph{semipermeable} membranes, permeable to small ions and solvent molecules, but not to large ions,  have been studied quantitatively during a century~\cite{donnan.fg:1924}. Such equilibria are of considerable importance for the theory of dialysis, as well as for the mechanism of cells and general physiology. Semipermeable membranes for modern industrial applications normally represent a polymer or ceramic film (at least several tens of micrometers thick)  with the array of nanopores or fibers~\cite{jong.j:2004,nikonenko.vv:2010}. However, there are many examples of a few nanometer thin semipermeable membranes, where the pore diameter is of the order of or larger than the membrane thickness. These are synthetic liposomes with ion channels~\cite{lindemann.m:2006}, stacked graphene oxide nanosheets~\cite{joshi.kr:2014,mi.b:2014}, multilayer shells of
polyelectrolyte microcapsules~\cite{donath.e:1998,vinogradova.oi:2006,vinogradova.oi:2004b,lulevich.vv:2004a}, or biological systems including viral capsids~\cite{cordova.a:2003,odijk.t:2003}, cell~\cite{alberts.b:1983,darnell.j:1986} and
bacterial~\cite{sen.k:1988,stock.jb:1977,sukharev.s:2001}
membranes. Since it was discovered, theories for interpreting Donnan equilibria mainly focussed on the case of a single infinitesimally thin membrane~\cite{bartlett.jh:1952,zhou.y:1988,deserno.m:2002}, or a single vesicle/capsule~\cite{zhou.y:1988,tsekov.r:2006,stukan.mr:2006,tsekov.r:2008,siber.a:2012}. However, ion equilibria can play a very important role in processes involving two membranes, such as adhesion~\cite{bruinsma.r:2000} or long-range electrostatic interactions between them~\cite{cowlew.ac:1978,kim.bs:2007}.

The quantitative understanding of electrostatic interactions of thin membranes is still challenging. Due to the tremendous complexity of real biological and synthetic membranes, the theory was restricted to very simple model membranes and relied on a number of assumptions and simplifications. Some solutions of the Poisson-Boltzmann equation (mean field theory) are known for charged bearing ionizable groups  immersed in the salt reservoirs~\cite{nimham.bw:1971}. The origin of a membrane charge is related to a dissociation of protons into the aqueous solutions, so that the focus of~\cite{nimham.bw:1971} was mainly on the role of the ionic strength and pH of the reservoir. The disjoining pressure was found to be repulsive with the essentially exponential long-range decay.  Later work assumed that the membrane is uncharged, and separated by a thin film of salt-free solvent from a charged wall~\cite{maduar.sr:2013} or another uncharged membrane~\cite{vinogradova.oi:2012}. Results were not limited by calculations within the Poisson-Boltzmann theory, and also included the Langevin dynamic simulations with explicit ions. The interaction with a wall was found to be repulsive and exponentially decaying~\cite{maduar.sr:2013}. However, for two interacting membranes  the power-law decay of a repulsive disjoning pressure was predicted in the large gap limit~\cite{vinogradova.oi:2012}. In both cases simulations confirmed the validity of the mean-field approach for a monovalent salt reservoir. Recent integral equation study of interactions of two vesicles immersed in an asymmetric electrolyte suggested that only in a case of low valency ions, interactions between vesicles are always repulsive. However, in large concentration solutions of trivalent co-ions  charge correlation effects  have been shown to result in short-range attractions between vesicles, so that only a long-range tail of interactions remain repulsive~\cite{lobaskin.v:2012}.

\begin{figure}
\centering
   \includegraphics[width=7cm]{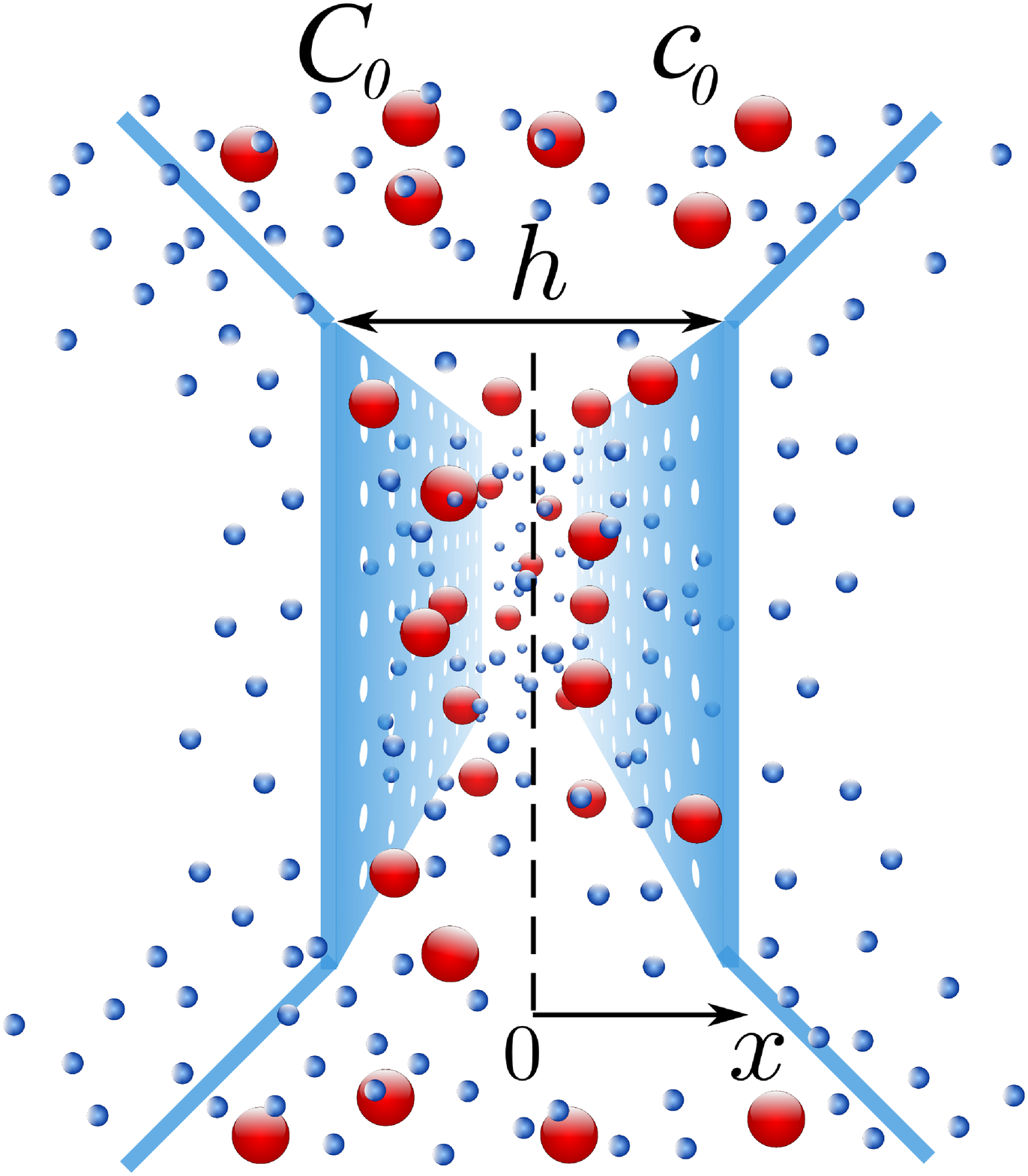}
  \caption{ Schematic illustration of the problem. An aqueous electrolyte solution is confined between two neutral semipermeable membranes, which are in contact with a salt-free solvent. Small ions leak-out to a salt-free reservoir, by giving rise to a surface potential and a repulsive disjoining pressure.}
  \label{fig:membranes}
\end{figure}

In this paper we study the electrostatic interaction of thin semipermeable membranes separated by aqueous electrolyte solutions. In other words, we address the problem, which is inverse to considered in recent work~\cite{vinogradova.oi:2012}. Such a configuration is relevant to interactions of large capsules/vesicles immersed in a salt reservoir, which can affect the aggregation properties of their suspension~\cite{kim.bs:2007,wittemann.a:2007}. We also note the relevance of our geometry to the problem of the adhesion of flattened shells immersed in water-electrolyte solutions~\cite{nardi.j:1998}. Our theoretical calculations will be limited  to solutions of Poisson-Boltzmann equation (mean-field theory). To verify the theory we will use Langevin dynamics simulations.

Our paper is organized as follows. Sec.2 contains a description of our model and methods, including our Langevin dynamics simulation approach. After some general considerations
of two flat membranes separated by an electrolyte solution and the formulation of the Poisson-Boltzmann equations, we present solutions for a distribution of a potential in the system within a non-linear  approach, and provide the Langevin dynamics simulations for results
which could potentially go beyond the continuum approach (Sec.3). Then, in Sec.4 we consider the osmotic pressure in the system, and relate it to a disjoining pressure in the gap.  Some concluding remarks are
presented in Sec.5. Appendix~\ref{DH} describes a linearized theory.

\section{Model and methods}

We consider two  flat  semipermeable membranes separated by a film of an electrolyte (salt) solution of a thickness $h$ as displayed in Fig.~\ref{fig:membranes}.  We restrict ourselves to a simplest case of uncharged membranes, which are treated in the continuum limit as a rigid infinitesimally thin interfaces. We will focus only on the long-range electrostatic properties of the system and do not consider van-der-Waals interactions, which obviously vanish in the limit of infinitesimally thin membrane. Let us recall that van-der-Waals forces in systems with finite membrane thickness have been studied before~\cite{tadmor.r:2001}. By applying this earlier analysis for a situation when the membrane thickness $d \ll h$, we can evaluate the van-der-Waals disjoining pressure, $\Pi \simeq - 2 A d^2/\pi h^5 \propto h^{-5}$ (here $A$ is the Hamaker constant), and conclude that it is of very short-range and much smaller values than expected for an electrostatic pressure.

The \emph{inner}, i.e. confined between membranes ($\vert x \vert< h/2$), solution contains large cations (or co-ions) of a charge $Z$ and small anion (or counter-ions) of a charge $z$.
This thin film is in a contact with a bulk reservoir of an electrolyte solution with concentrations of large ions, $C_0$, and of small ions, $c_0$, which satisfies an electroneutrality condition, $z e\ c_0 + Z e\ C_0=0 $.
Here we should like to emphasize the key difference from the earlier considered problem of a shell filled with an electrolyte solution, where the number of large ions in the confined spherical volume was fixed~\cite{tsekov.r:2006}.  Coming back to the formulation of our model, we assume that cations cannot permeate through membranes, while anions (due to entropy reasons) can leak-out into \emph{outer} ($\vert x \vert> h/2$) infinite reservoirs of a solvent. As a result, we get a steric charge separation: an inner solution becomes positively, and an outer solution negatively charged. As any charged object it attracts a cloud of counter-ions of the outer area forming an outer diffuse electric double layer. Note that since no electrolyte is added to the solvent, and the only ions in the outer solution are the leaked-out counterions balancing exactly the excess charge in the film.

The theoretical calculations will be limited to
solutions of the Poisson-Boltzmann equation (mean-field theory) and its linearized version. In such a description, the finite
size of the ions and correlations are also ignored.  The
electric potentials and concentrations of ions are described by
continuous variables (using the continuum hypothesis).


The thermodynamic equilibrium of ions in the inner and outer regions with the bulk reservoir leads to the Boltzmann distributions, with the reference concentrations ($c_0,C_0$) that correspond to a bulk solution where the potential is equal to zero:
 \begin{eqnarray}\label{boltzmann}
c_{i,o}(x)=c_0 \exp[-\varphi_{i,o}(x)]  \\
C_{i}(x)=C_0 \exp[-\tilde Z \varphi_{i}(x)].
 \end{eqnarray}
Here
\begin{equation}
\varphi_{i,o}={z e\phi_{i,o}\over k_BT},
\end{equation}
are the dimensionless electrostatic potentials, the indices $\{i,o\}$ indicate inner and outer solutions, and $\tilde Z=Z/z$ ($<0$) the valence ratio of large and small ions.

For results which could potentially go beyond the continuum approach, we employ the Langevin dynamics (MD) simulations with explicit ions, by using
the ESPResSo simulation package~\cite{Espresso}. We follow the approach developed earlier~\cite{vinogradova.oi:2012}, so that we only recall now the basic ideas and parameters of the model.
A more detailed description of the basic steps of this approach and some additional simulation details will be given in Sec.3 and Sec.4.

Ions were implemented as Lennard-Jones spherical particles with a central charge. The charge of counter-ions was always fixed as $z = -1$, but we varied the charge of large ions from $Z = 1$ to $5$. Repulsive inter-ionic Lennard-Jones potential was set with cut-off distance $r_c=2^{1/6}\sigma$:
\begin{eqnarray}
U_{LJ}(r)=
\begin{cases}
4\epsilon\left[\left(\frac{\displaystyle\sigma}{\displaystyle r}\right)^{12}-\left(\frac{\displaystyle
\sigma}{\displaystyle r}\right)^6 +\frac{\displaystyle 1}{\displaystyle 4} \right], \quad r\leq r_c; \\
0,\quad r>r_c,
\end{cases}
\label{potlj_txt}
\end{eqnarray}
where $r$ is the distance between centers of two particles. The
energy parameter $\epsilon$ was fixed to be equal to $k_B T$, and the particle size $\sigma$ was equal to unity for all types of interactions. The solvent was treated as a homogeneous medium with the Bjerrum length
\begin{equation}\label{Bjerrum}
  \ell_B=\frac{e^2}{4\pi\epsilon\epsilon_0k_BT},
\end{equation}
which varied in the interval from $0.4\sigma$ to $\sigma$. For aqueous solutions, $\ell_B=0.7$ nm, which implies that $\sigma$ is of the order of $0.7-1.75$ nm.

The interaction of ions with membranes was set by
\begin{eqnarray}
U_{LJ}(x)=
\begin{cases}
4\epsilon\left[\left(\frac{\displaystyle\sigma}{
\displaystyle x}\right)^{12}-\left(\frac{\displaystyle
\sigma}{\displaystyle x}\right)^6 -\left(\frac{
\displaystyle\sigma}{\displaystyle x_c}\right)^{12}+\left(\frac{
\displaystyle\sigma}{\displaystyle x_c}\right)^6\right],\\ \quad  x \leq x_c; \\
0,\quad x>x_c,
\end{cases}
\label{potlj_2}
\end{eqnarray}
with the cut-off distance $x_c=2^{1/6} \sigma$.

The electrostatic
interaction between ions was modeled by the Coulomb
potential implemented in 3D rectangular periodic cell ($L_x, L_y, L_z$), which should be large enough to provide a nearly vanishing concentration of small ions at large $x$. It was computed by using the P3M algorithm~\cite{P3M1989} with maximum
relative accuracy of $10^{-5}$. In our simulations we confined in the gap from 1200 to 4000 ions to reach sufficient statistical accuracy and to provide required concentrations.

\section{Potentials}

We begin by studying a distribution of electrostatic potentials in our system. Below we present theoretical analysis and example calculations based on the continuous theory, as well as the results of computer simulations.

\subsection{Poisson-Boltzmann equations and formal solutions. }


The non-linear Poisson-Boltzmann equations (NLPB) for  the dimensionless electrostatic potentials
take the form:
\begin{eqnarray}
& \dfrac{\partial^2 \varphi_i(x)}{\partial x^2}&= - \kappa_o^2 \left( e^{-\varphi_i}-e^{-\tilde Z \varphi_i} \right) \label{NLPB-1}
\\
& \dfrac{\partial^2 \varphi_o(x)}{\partial x^2}&= - \kappa_o^2\, e^{-\varphi_o}
\label{NLPB-2}
\end{eqnarray}
with an outer inverse screening length
\begin{equation}\label{kappao}
\kappa_o^2=4\pi  \ell_B z^2 c_0
\end{equation}
 Due to electroneutrality of the system, an inner inverse screening length is related to an outer one as
 \begin{equation}\label{kappai}
\kappa_i^2=\kappa_o^2 (1 - \tilde Z)
\end{equation}
 We remark and stress that Eq.(\ref{NLPB-1}) does not use earlier assumption that counter-ions will leak out from the confined volume  completely~\cite{tsekov.r:2006}.  In contrast, it allows one counter-ions remain inside the gap.

Eq.(\ref{NLPB-2}) can be integrated once to give
\begin{equation}
{1\over 2\kappa_o^2} \left({\partial \varphi_o\over \partial x}\right)^2=e^{-\varphi_o}+B,
\label{phi-out-2}
\end{equation}
where the integration constant $B$ is determined by the boundary condition at infinity. Due to the absence of large ions in the outer space, one expects the potential to go to infinity
as $\vert x\vert \rightarrow \infty$. We seek for a solution verifying also $\partial_x\varphi(x\rightarrow\infty)=0$, {\it i.e.} $B=0$.

 In this case, integration of Eq. \eqref{phi-out-2} gives
\begin{equation}
\varphi_o(x)= 2 \ln\left[ e^{\varphi_s/2}+ {\kappa_o\over \sqrt{2}}\left(x-{h\over 2}\right)\right]\label{x-out-2}
\end{equation}
Lets remark that the large $x$ behavior, $\varphi_o(x) \simeq \ln(x)$ (note the similarity with the classical Gouy-Chapman theory of charged surfaces in a salt-free solvent~\cite{Andelman_biophysics}), does indeed verify $\partial_x\varphi(x\rightarrow\infty)\simeq x^{-1}\rightarrow0$.

Similarly, Eq.(\ref{NLPB-1}) can be integrated once by using  $\partial_x \varphi(x=0)=0$ to give:
\begin{equation}
{1\over 2\kappa_o^2} \left({\partial \varphi_i\over \partial x}\right)^2=e^{-\varphi_i}-{1\over\tilde Z} e^{-\tilde Z \varphi_i}-\gamma_m
\label{dphi-o-2}
\end{equation}
where
\begin{equation}
 \gamma_m= p_m/k_BTc_0 = e^{-\varphi_m}-{1\over\tilde Z} e^{-\tilde Z \varphi_m}
\end{equation}
is the normalized osmotic pressure in the middle of the gap, expressed solely in terms of the mid-plane potential $\varphi_m$. Thus, the expression for $\varphi_i$ takes the form:

\begin{equation}
\int_{\varphi_m}^{\varphi_i}{d\varphi \over \sqrt{2\left( \exp[-\varphi]-{1\over\tilde Z} \exp[-\tilde Z\varphi ]-\gamma_m\right)}}=- \kappa_o\,x
\label{xin-2}
\end{equation}

The membrane, $\varphi_s$, and mid-plane, $\varphi_m$, potentials are then given by the self-consistency equations

\begin{equation}
\int_{\varphi_m}^{\varphi_s}{-d\varphi \over \sqrt{2\left( \exp[-\varphi]-{1\over\tilde Z} \exp[-\tilde Z\varphi ]-\gamma_m\right)}}= \kappa_o\, {h\over 2}
\label{SC-1-1}
\end{equation}
and
\begin{equation}
\gamma_m=-{e^{-\tilde Z \varphi_s}\over\tilde Z}
\label{SC-2-1}
\end{equation}

Note that Eq.~(\ref{SC-2-1}) is obtained by subtracting Eq.~\eqref{phi-out-2} from Eq.~\eqref{dphi-o-2} at $x=\pm h/2$ with subsequent using a condition of the continuity of the electric field across the neutral membrane.

\begin{figure}[h]
\centering
\includegraphics [width=8.5cm]{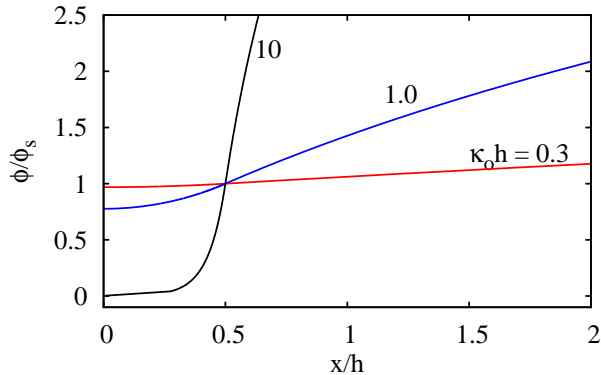}
  \caption{A distribution of a potential at $\kappa_o h = 0.3, 1,$ and $10$, normalized
by a surface potential, computed within the NLPB theory, $\tilde{Z}=-1$.}
  \label{fig:NLPB:pot_distrib}
\end{figure}

To obtain $\varphi_s$ and $\varphi_m$ Eqs.~(\ref{SC-1-1}-\ref{SC-2-1}) should be resolved numerically at fixed $\kappa_o h$. This allows one to calculate then the spatial distribution of an electrostatic potential in the system. The distribution of a potential computed for different
$\kappa_o h$ is shown in Fig.~\ref{fig:NLPB:pot_distrib}.  All curves are normalized
to the corresponding value of $\varphi_s$. It can be seen that the
potential diverges at infinity and takes the minimum value at the center of the gap, and that its value  depends strongly on $\kappa_o h$.

\begin{figure}[h]
\centering
\includegraphics[width=8.5cm]{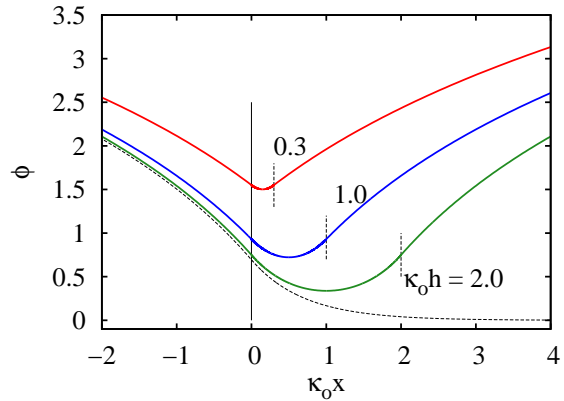}
\caption{The effect of  a distance between membranes on a distribution of a potential in the system predicted by the NLPB theory, $\tilde{Z}=-1$. One membrane is located at $x=0$. Another membrane is at a distance $\kappa_o h =0.3, 1,$ and $2$ (solid curves from top to bottom). Vertical lines show the locations of membranes. Dashed curve corresponds to a situation of an isolated membrane. }\label{membrane_effect}
\end{figure}

We can now compare the distribution of the potential in the system depending on a separation between membranes (see Fig.~\ref{membrane_effect}). We start by considering an isolated membrane located at $x=0$, which will be a reference system for our consideration. The potential decays to zero at large positive $x$, but logarithmically diverges when $x$ is large and negative. Obviously, at the membrane surface it takes the value of the bulk Donnan potential~\cite{vinogradova.oi:2012}
\begin{equation}
\varphi_s \simeq - \frac{\ln (1-\tilde{Z})}{\tilde{Z}}\label{surface_pot_ls}
\end{equation}
  If second membrane is fixed at some finite $x$, the absolute value of a potential increases with a decrease in separation between membranes. The potential shows  a minimum at the midplane, but diverges in the outer regions far from membranes.

  \begin{figure}[h]
\centering
\includegraphics [width=8.5cm]{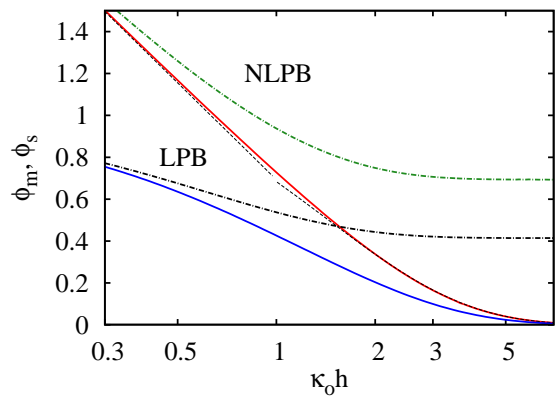}
  \caption{Surface (dash-dotted curves) and midplane (solid curves) potentials versus $\kappa_o h$ computed for $\tilde{Z} = -1$. Upper curves correspond to NLPB, and lower curves to LPB theory. Dashed lines show predictions of asymptotic Eqs.~(\ref{NLPB_phim_ss}) and (\ref{NLPB_phim_ls}).}
  \label{fig:NLPB:psi_m_psi_s}
\end{figure}

Electrostatic potentials, $\varphi_s$ and $\varphi_m$, versus $\kappa_o h$ predicted by the NLPB theory are shown in Fig.~\ref{fig:NLPB:psi_m_psi_s}. This plot illustrates that $\varphi_s$ and $\varphi_m$ diverges at small $\kappa_o h$, which is caused by a divergence of the integrand in Eq.~\eqref{SC-1-1}. By using Eq.~\eqref{SC-2-1} and that in this limit $\varphi_s-\varphi_m \ll \varphi_s$ one can derive
\begin{equation}\label{NLPB_phim_ss}
\varphi_m \simeq \varphi_s \simeq \frac{2}{2\tilde{Z}-1}\ln\left(\frac{\kappa_o h}{2\sqrt{2}}\right)
\end{equation}
This first-order asymptotic result is included in  Fig.~\ref{fig:NLPB:psi_m_psi_s}, and we see that it is in good agreement with the numerical calculations. Fig.~\ref{fig:NLPB:psi_m_psi_s} also shows that $\varphi_s$ asymptotically approaches the constant (at fixed $\tilde{Z}$) Donnan potential, given by Eq.(\ref{surface_pot_ls}), and that $\varphi_m$ vanishes at large $\kappa_o h$.

By assuming small potentials $\varphi \ll 1$ at $\kappa_o h \gg 1$, we may linearize the integrand in Eq.~(\ref{SC-1-1})
to derive
\begin{equation}
\varphi_m \simeq 2 \varphi_s \exp \left(-{\kappa_i h\over 2}\right). \label{NLPB_phim_ls}
\end{equation}
Eq.(\ref{NLPB_phim_ls}) indicates that the midplane potential decays exponentially with a separation, and that the decay length is equal to $2 \kappa_i^{-1}$. Substituting in Eq.(\ref{NLPB_phim_ls}) the expression for a surface potential of an isolated membrane, Eq.(\ref{surface_pot_ls}), we get
\begin{equation}
\varphi_m \simeq - \frac{2 \ln (1-\tilde{Z})}{\tilde{Z}} \exp \left(-{\kappa_i h\over 2}\right)
\end{equation}
A key remark is that the assumption of a small potential used to derive Eq.(\ref{NLPB_phim_ls}) is not really justified since the surface potential given by Eq.(\ref{surface_pot_ls}) is not small enough. However, predictions of Eq.(\ref{NLPB_phim_ls}) are
in excellent agreement with the numerical results as seen in Fig.~\ref{fig:NLPB:psi_m_psi_s}.

An important remark is that although membranes are neutral one can introduce effective surface charge density (at the imaginary impermeable solid surface, which mimics the actual membrane by  generating the same distribution of a potential in the inner region):

\begin{equation}\label{eff_charge}
    \frac{q}{e}=\frac{\partial_x\varphi}{4\pi z\ell_B} = \dfrac{1}{2\sqrt2 \pi}\frac{e^{-\varphi_s(h)/2}}{z \ell_B \kappa_o^{-1}}
\end{equation}
Note that in contrast to conventional impermeable solids, the effective charge density of membranes depends on separation between them and on the screening length. One can suggest that if the effective surface charge density becomes high, and there are high valency ions in the electrolyte solution, charge correlations and
charge fluctuations could become important and the NLPB approach could fail. The justification of the NLPB approach in our system of two membranes will be clear \emph{a posteriori}, and we will return to this issue below.

Finally, we recall that at low values of the electric potential, the description of our problem can be simplified by linearization of the Poisson-Boltzmann (LPB) approach, which is discussed in Appendix~\ref{DH}. Theoretical curves calculated with LPB theory are included in Fig.~\ref{fig:NLPB:psi_m_psi_s}.  Except the  decay  to zero of the midplane potential at large $\kappa_o h$, which is similar to $ \varphi_m$ in NLPB, there is a discrepancy between LPB and numerical results for both potentials. The discrepancy is always in the direction of the smaller potentials than predicted by the NLPB. In the limit of large $\kappa_i h$ the surface potential is equal to a Donnan potential in the LPB theory~\cite{tsekov.r:2008}
\begin{equation}\label{donnan_LPB}
  \varphi_s=\frac{1}{1+\sqrt{1-\tilde{Z}}},
\end{equation}
which value differs from the Donnan potential in the NLPB approach, Eq.(\ref{surface_pot_ls}). For small $\kappa_i h$ LPB theory predicts $\varphi_s\simeq \varphi_m\simeq 1$, i.e. in contrast to the NLPB theory, the surface and midplane potentials in the LPB approach do not diverge. This result is similar to obtained earlier for an isolated semipermeable shell immersed in electrolyte solutions~\cite{lobaskin.v:2012}. Altogether, for our system the LPB theory is very approximate, but it provides us with some guidance.

\subsection{Computer Simulation}

\begin{figure}
\begin{center}
\includegraphics [width=8.5 cm]
{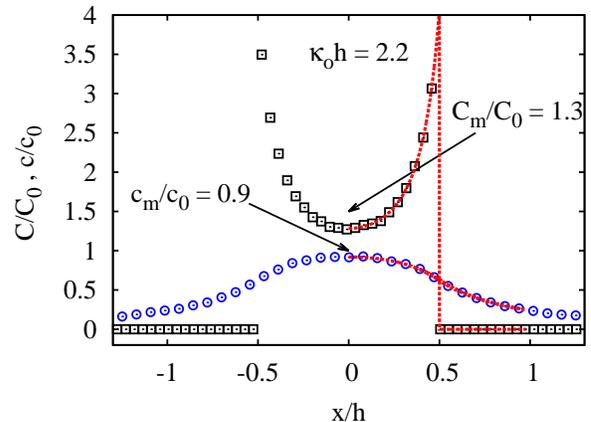}
   \end{center}
  \caption{Simulated concentration profiles of large and small ions (symbols) and results of the NLPB theory (dashed curves). Parameters are $\tilde{Z} = -3$, $\kappa_o h=2.22$. }
  \label{fig:simulation_conc}
\end{figure}

In order to assess the validity of the above NLPB mean-field approach, which ignores correlations and/or finite size effects that can be important for multivalent ions and large concentrations, we perform Langevin dynamics (MD) simulations.

  In our simulations we used ${L}_y, L_z \geq 150\sigma_{LJ}$, and $L_x$ was always larger than $\simeq 30 \kappa_o^{-1}$. With our parameters we were able to vary the outer screening length in the interval of $\kappa_o^{-1} \simeq (20-50) \sigma$.

Initial positions of ions were set randomly in the gap between membranes. During the equilibration step small ions leak out to the outer space. After the equilibrium was reached we measured ion concentration profiles. Since in our simulation setup the bulk reservoir with an electrolyte solution was absent, we evaluated $C_0, c_0$ a posteriori from measured local concentrations of ions by
using the Boltzmann distribution, Eq.(\ref{boltzmann}).

The detailed comparison between the simulation results and the NLPB theory is then shown in Fig.~\ref{fig:simulation_conc}. It can be seen that a quantitative agreement is reached even for this example simulations with multivalent ions, $\tilde{Z} = -3$. We have performed similar calculations for several $c_0$ in the range  $(10^{-4} - 10^{-3}) \sigma^{-3}$, which for water systems would correspond to $c_0$ varying from $3\times10^{-5}$ to $5\times10^{-3}$ mol/L. In all cases we found an excellent agreement with the NLPB predictions, suggesting that in this range of parameters the mean-field theory can safely be used.

For charged impermeable surfaces the charge density, $q$, the NLPB approach is violated in the strong-coupling regime~\cite{grosberg.ay:2002,levin.y:2002,netz.rr:2001}, where the so-called electrostatic coupling parameter~\cite{moreira.ag:2002,Kanduc.M:2008}, $\Sigma=2\pi \vert Z\vert^3 \ell_B^2 \left|\dfrac{q}{e}\right|$ is large enough. For all separations between our membranes $\Sigma \leq 2$. By using the criteria proposed earlier for counter-ions near a single charged surface~\cite{netz.rr:2001}, we can conclude that with our concentrations and valency of counter-ions the correlations in the outer solutions are generally impossible. The same conclusion is true for an inner solution. Note that in the inner solution deviations from the NLPB can also appear due to a formation of ionic pairs~\cite{dossantos.ap:2010}, which is also negligibly small with our concentrations.

\begin{figure}
\begin{center}
\includegraphics[width=8.5cm]{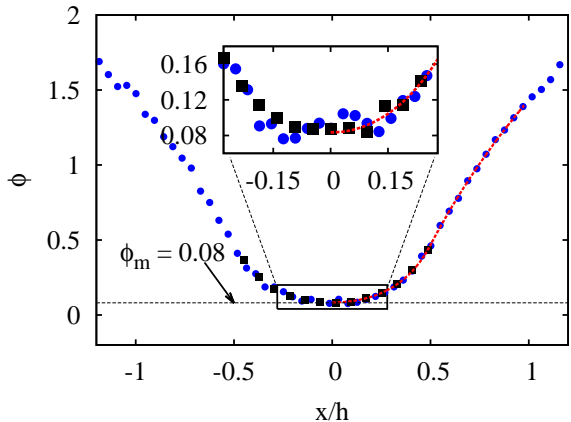}
\end{center}
\caption{Typical distribution of a potential calculated from simulated small (circles) and large (rectangles) ion concentration profiles, $\kappa_o h=2.22;\  \tilde{Z}=-3$. Dashed curve shows results of a numerical solution of NLPB equations.}
\label{Fig:simulation_potential}
\end{figure}

Electrostatic potential was then deduced from measured concentration profiles by using Boltzmann distributions, Eq.~\eqref{boltzmann},
and results were again validated by comparison with theoretical NLPB predictions. The data presented in Fig.~\ref{fig:simulation_conc} are therefore now used in Fig.~\ref{Fig:simulation_potential} to plot the distribution of a potential in our system. A key remark is that a potential calculated by using concentration profiles of small and large ions coincide, confirming the validity of our approach.

\underline{}

\section{Pressure}

In this Section we consider the relationship between osmotic (and relevant disjoining) pressure and parameters of the system, such as a separation between membranes, and parameters of an electrolyte solution.

\subsection{Non-linear theory. }

We first write the force balance in the inner and outer regions
\begin{equation}
-\nabla p + \rho_c E =0 \label{hydrostatic}
\end{equation}
with $p$, an osmotic pressure,  $\rho_c$ the charge density, and $E=-\partial_x \phi$ the local electric field. Using the Boltzmann expressions for the charge densities in terms of the local electrostatic potentials allows to
integrate this equation once.

In the inner region this leads to
\begin{equation}
p_{i}(x)=k_BT c(x) + k_BT C(x) + p_0
\label{pin}
\end{equation}
with $p_0$ a constant.

In the outer space one gets
\begin{equation}
p_{o}(x)=k_BT c(x) + p_\infty
\label{pout}
\end{equation}
with  $p_\infty$ the
value of the pressure at infinity.

At the membrane, there is a pressure drop proportional to the difference of large ions concentration on the two sides of the membrane, {\it i.e.}:
\begin{equation}
\Delta p = p_{i} ( {h\over 2}^+)-p_{o}({h\over 2}^-)=k_BT C({h\over 2}^-)
\label{DP}
\end{equation}
This imposes $p_0=p_\infty$, the solvent pressure, as expected.

 The disjoining pressure, $\Pi$, is defined as:
 \begin{equation}\label{disj_definition}
\Pi=\Delta p - p_{id},
\end{equation}
where $p_{id} = k_BT(c_0+C_0)$, so that we can express it through the surface potential
\begin{equation}
\Pi=k_BTC_0 e^{-\tilde{Z}\varphi_s} - p_{id},
\end{equation}
and then, by using Eq.~\eqref{SC-2-1}, through the midplane potential:
\begin{equation}
\Pi=k_BT c_0 \left\{ e^{-\varphi_m}-1 -\frac{1}{\tilde{Z}} \left(e^{-\tilde Z\varphi_m}-1\right)\right\}\label{NLPB:disj_press}
\end{equation}

\begin{figure}[h]
\centering
\includegraphics [width=8.7 cm]{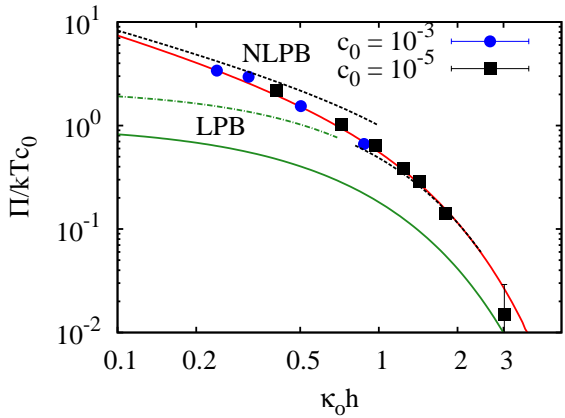}
     \caption{Disjoining pressure versus $\kappa_o h$ computed for monovalent ions, $\tilde{Z} = -1$. Solid curves correspond to calculations with NLPB and LPB theories. Symbols show  simulation results for two different concentrations of small ions in bulk electrolyte solution. Dashed lines are calculations with asymptotic formulae, Eqs.~(\ref{pressure_ss}) and (\ref{eq:disj_asymp_ls}). Dash-dotted curve shows a LPB solution obtained by substitution of the NLPB membrane potential, Eqs.~(\ref{surface_pot_ls}).}
  \label{fig:NLPB:Disj}
\end{figure}

Fig.~\ref{fig:NLPB:Disj} include theoretical curves calculated with Eq.~\eqref{NLPB:disj_press}, by using NLPB values of $\varphi_s$ and $\varphi_m$. By using Eqs.~(\ref{NLPB_phim_ss}) and (\ref{NLPB_phim_ls}) one can also construct  asymptotic expressions.

For large $\kappa_o h$ the disjoining pressure decays exponentially

\begin{equation}
\Pi \simeq 2 {kTc_0} (1-\tilde{Z})\varphi_s^2 \exp(-\kappa_i h),
\label{eq:disj_asymp_ls}
\end{equation}
where (constant) $\varphi_s$ is given by Eq.~\eqref{surface_pot_ls}. Similar asymptotic decay is typical for an interaction of two charged impermeable~\cite{Andelman:2010_PB_revisit} and ion-permeable porous~\cite{Ohshima2010:planes} surfaces.  Note that some prior work~\cite{tsekov.r:2006} predicted a power-law divergence of $\Pi$ at large distances, which is likely the consequence of an unrealistic initial assumption that all counter-ions leave the confined volume.

 In the limit of small $\kappa_o h$ Eq.(\ref{NLPB:disj_press}) can be simplified to $\Pi \simeq k_BT C_0 e^{-\tilde Z \varphi_m}$. By using Eq.\eqref{NLPB_phim_ss} we then get
 \begin{equation}
 \Pi \simeq \frac{k_BT C_0}{ (\kappa_o h)^{\alpha}},\label{pressure_ss}
 \end{equation}
 with an exponent $\alpha= {2 \tilde Z \over 2\tilde Z -1}$, which depends only on $\tilde Z$ and takes values from $2/3$ (at $\tilde{Z}=-1$) to  $\simeq 1$ (for polyions). This conclusion generalizes earlier results on a power-law decay of a disjoining pressure obtained for charged impermeable surfaces~\cite{McCormack1995:PB_solution_2plates} and neutral uncharged membranes~\cite{vinogradova.oi:2012}.  These include a repulsion of strongly charged surfaces~\cite{Israelachvili2011, Nato_andelman} (Gouy-Chapman regime) and membranes in contact with bulk solutions separated by a thin film of salt-free liquid~\cite{vinogradova.oi:2012}, where a disjoining pressure decays as $\Pi \propto h^{-2}$. An interaction of weakly charged surfaces in the so-called ideal gas regime scales as $\Pi \propto h^{-1}$~\cite{Nato_andelman, Andelman_biophysics}.

Fig.~\ref{fig:NLPB:Disj} also includes the theoretical curves
calculated within LPB theory (see Appendix~\ref{DH}). We see that LPB theory significantly underestimates the value of $\Pi$. Note however that by using the bulk NLPB Donnan potential, Eq.(\ref{surface_pot_ls}), instead of Eq.(\ref{donnan_LPB}), to calculate LPB disjoining pressure, we obtain a good agreement between LPB and NLPB results for large $\kappa_o h$. So, in this regime the discrepancy is obviously only due to different values of Donnan potentials in the NLPB and LPB theories.

\begin{figure}[h]
\centering
\includegraphics [width=8.5cm]{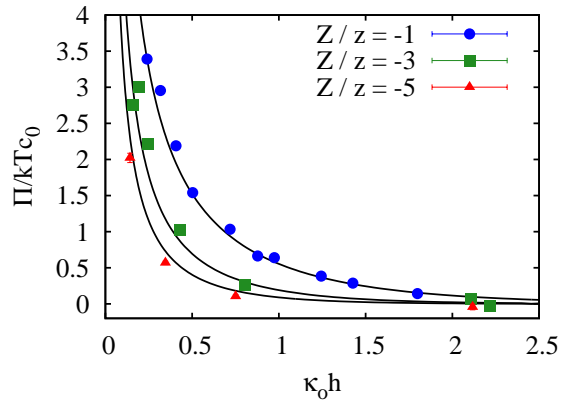}
    \caption{The disjoining pressure versus $\kappa_o h$ calculated by using NLPB approach. Solid lines from top to bottom correspond to $\tilde{Z} = -1, -3$ and $-5$. Symbols show simulation data.}
  \label{fig:simulation_Disj_multion}
\end{figure}

Finally, we consider in this paragraph the effect of multivalent ions on a disjoining pressure. For our system they are expected to influence screening lengths, as well as to decrease the surface and midplane potentials. Fig.~\ref{fig:simulation_Disj_multion} includes theoretical curves for $\Pi$ as a function of $\kappa_o h$ calculated for different $\tilde{Z}$ by using NLPB theory. It can be seen that the increase in the absolute value of $\tilde{Z}$ reduces the range and strength of the repulsive interaction between membranes.

\subsection{Comparison with computer simulations. }

Using computer simulations we can now verify the above expressions for a disjoining pressure following the earlier developed approach~\cite{vinogradova.oi:2012}. Briefly, pressure has been evaluated via integration of the LJ force of
cations, acting on the membrane walls
\begin{equation}\label{fx}
F(x) = 4 \varepsilon \left(\frac{12\sigma^{12}}{\left(x-\frac{h}{2}\right)^{13}}-\frac{6\sigma^{6}}
{\left(x-\frac{h}{2}\right)^{7}}\right),
\end{equation}
which gives the pressure drop at the membrane
\begin{equation}
\Delta p=\int_{h/2}^{h+2^{1/6} \sigma} C(x) F(x) dx.
\end{equation}
The latter was averaged over more than 50000 independent simulations during $5\times 10^6 \tau$, where $\tau=\sigma\sqrt{m/\epsilon}$ is the characteristic time scale in our system. Using measured values of $C_0, c_0$ and Eq.~\eqref{disj_definition} we then calculated the disjoining pressure in computer simulation.

We have simulated the disjoining pressure at different $\kappa_o h$ and included results in Fig.~\ref{fig:NLPB:Disj} for comparison with theoretical predictions. A general conclusion is that the predictions of the NLPB theory, including asymptotic results, in the previous
section are in excellent agreement with simulation data.
Altogether the simulation results do confirm the mean-field predictions with our concentrations of ions. So there is no indication of an attractive force that could  appear in case of more concentrated solutions of multivalent ions~\cite{Kanduc.M:2008,Naji2005,lobaskin.v:2012}.

\section{Concluding remarks}

In conclusion, we developed a mean-field theory of electrostatic interactions of two semipermeable membranes separated by a thin film of an aqueous electrolyte solution.
 The theory allows to provide explicit analytical expressions for the disjoining pressure in some asymptotic limits. We have predicted power-low decay of a disjoining pressure with the separation between membranes at short distances, and an exponentially decaying disjoining pressure in case of thick gap.
  We checked the validity of this approach by Langevin dynamics simulations.  The simulation results show that a non-linear theoretical description remains valid even for multivalent ions. However, only a qualitative agreement can be obtained by using the linearized theory.

Finally, we mention that all results found in the present study for neutral membranes and salt-free outer reservoirs can be immediately
extended for a more complex case of charged membranes and added salts. As an extension of this study, our results could also be applied in a different field to explore the questions
of conformations of membrane proteins, ionic channels, signalling between cells caused by changes in membrane potentials due to various interactions~\cite{Gingell1967:surface_potential_cell_interaction,
Honig:1986}.

\section*{Acknowledgements}
We have benefited from discussions with L.~Bocquet and R.~Tsekov
at the initial stage of this study.
This work was supported by the Russian Foundation for Basic Research (grant 12-03-00916). Access to
computational resources at the Center for Parallel Computing
at the M.V. Lomonosov Moscow State University
(``Lomonosov'' and ``Chebyshev'' supercomputers) is gratefully
acknowledged.

\appendix
\section{Linearized theory}\label{DH}

Here, we propose a simpler description, which is possible in the limit of small potentials, where Eqs.~\eqref{NLPB-1} and \eqref{NLPB-2} can be linearized  (LPB):
\begin{eqnarray}
& \dfrac{\partial^2 \phi_i(x)}{\partial x^2} &\simeq \kappa_i^2 \varphi_i \label{PE_in_DH1} \\
& \dfrac{\partial^2 \phi_o(x)}{\partial x^2} &\simeq \kappa_o^2 (\varphi_o-1) \label{PE_in_DH2}
\end{eqnarray}

The inner and outer solutions are then
\begin{equation}
 \varphi_i(x)\simeq\frac{\cosh(\kappa_i x)}{\cosh(\kappa_i h/2)}\varphi_s
 \end{equation}
 \begin{equation}
 \varphi_o(x)\simeq1+(\varphi_s-1) e^{\kappa_o(h/2-x)}
\end{equation}
with
\begin{equation}\label{varphis_LPB}
  \varphi_s\simeq\frac{1}{1 + \sqrt{1-\tilde{Z}} \tanh\left(\kappa_i h  /2\right)}
\end{equation}

The midplane and surface potentials are related by simple formula
\begin{equation}\label{varphim_LPB}
 \varphi_m\simeq\frac{\varphi_s}{\cosh(\kappa_i h/2)}
 \end{equation}

Coming back to the LPB approach a similar analysis based on a solution of Eq.~\eqref{hydrostatic} together with Eqs. (\ref{PE_in_DH1}) and (\ref{PE_in_DH2}) gives
\begin{equation}
\Delta p \simeq p_{id}+{k T c_0} \left\{\varphi_s-\frac{{\tilde{Z}}\varphi_s^2}{2}-\frac{1}{2}\right\},
\end{equation}
so that for the disjoining pressure we get
\begin{equation}\label{Pi_LPB}
\Pi \simeq {k T c_0} \left\{\varphi_s-\frac{{\tilde{Z}}\varphi_s^2}{2}-\frac{1}{2}\right\}
\end{equation}
We emphasize that despite an original assumption of potentials, which are small enough to
take their linear contribution to the ion distribution in the system,  Eq.~\eqref{Pi_LPB} is nonlinear and contains the quadratic $\varphi_s$-term to provide a self-consistency of the LPB theory. A discussion  of physical ideas underlying routes to calculate the pressure in the LPB can be found in~\cite{deserno.m:2002,tsekov.r:2008}.

Eq.(\ref{Pi_LPB}) can be rewritten as
\begin{equation}
\Pi = k T c_0 \frac{(1-\tilde{Z})\varphi_m^2}{2}, \label{PE_in_disj}
\end{equation}
where $\varphi_m$ is given by Eq.(\ref{varphim_LPB}).

At $\kappa_o h \gg 1$ we can derive the same (exponential) asymptotics as in the NLPB theory, Eq.(\ref{eq:disj_asymp_ls}), but now (constant) $\varphi_s$ is given by Eq.~\eqref{varphis_LPB}.
At $\kappa_o h \ll 1$, $\varphi_s \simeq \varphi_m \simeq 1$, and we get:
\begin{equation}
\Pi \simeq k T c_0 \frac{2(1-\tilde{Z})}{\left(2+ \kappa_o h(1-\tilde{Z})\right)^2},
\end{equation}
i.e. we predicted a power-law decay, and finite $\Pi \simeq k T c_0 (1-\tilde{Z})/2$ at a contact.


\end{document}